\documentclass[a4paper,11pt,twocolumn]{article}
\usepackage{geometry}
 \usepackage{hyperref}	
 \hypersetup{colorlinks=true,citecolor=red}
\usepackage{subcaption}
\usepackage[utf8]{inputenc} 
\usepackage[T1]{fontenc} 
\usepackage{mathpazo} 
\usepackage{tikz}
\usetikzlibrary{positioning,decorations.pathreplacing,shapes}
\usepackage{graphicx} 
\usepackage{microtype}
\usepackage{booktabs} 
\usepackage{listings} 
\usepackage{enumerate} 
\usepackage{amsmath,bm}
\usepackage{authblk}
\usepackage{amsfonts}
\usepackage{amssymb}
\usepackage{braket}
\usepackage{mathrsfs}
\usepackage{color}
\usepackage{simpler-wick}
\usepackage{physics}
\usepackage{cite}
\usetikzlibrary{graphs}

\usepackage{hyperref}	
\hypersetup{colorlinks=true,citecolor=red}

\usepackage{dcolumn}
\usepackage{bm}

\usepackage{mathptmx}
\usepackage{etoolbox}
\usepackage{subcaption}

\usepackage{soul}

\begin{document}

\author[a]{ Zexiang Deng$^*$}

\affil[a]{School of Science, Guilin University of Aerospace Technology, Guilin 541004, People's Republic of China}

\title{Prediction on Raman Spectra of Intrinsic Two-Dimensional Ga$_2$O$_3$ Monolayer }

\twocolumn[

\maketitle

\begin{abstract}
	We investigate the vibrational properties and Raman spectra of two-dimensional Ga$_2$O$_3$ monolayer, using density functional theory. 
	Two ferroelectric (FE) phases of Ga$_2$O$_3$ monolayer with  wurtzite(WZ) and zincblende(ZB) structures( FE-WZ and FE-ZB, respectively ) are considered.	
	The Raman tensor and angle-dependent Raman intensities of two major Raman peak ($A^1_1$ and $A^2_1$) in both FE-WZ (497, and 779 cm$^{-1}$) and FE-ZB  (481, and 772 cm$^{-1}$)  Ga$_2$O$_3$ monolayers are calculated for the polarizations of scattered light parallel and perpendicular to that of the incident light. The characteristics of angle-dependent Raman intensities are analyzed. 
	The averaged non-resonant Raman spectra of minor peaks in FE-WZ($E^1$) and FE-BZ($E^1$ and $E^2$) are compared with that of major peaks $A^1_1$ and $A^2_1$ . These predictions on Raman spectra of Ga$_2$O$_3$ monolayer may guide rational design of two-dimensional optical devices.
	
\end{abstract}

\bigskip


\bigskip

\bigskip

]

 \section{Introduction}
 Materials such as ZnO, GaN, AlN, Ga$_2$O$_3$\cite{tsao2018ultrawide} have attracted the attention of many research groups due to their novel wide band gaps.
 Low-dimensional nanostructure ultraviolet photodetectors(ultrawide band gap up to 4.8eV \cite{pearton2018review,higashiwaki2016recent,kaur2021strategic}) are widely investigated.\cite{peng2013low} Ultrahigh performance of Ga$_2$O$_3$ (band gap = 4.6-5.3 eV) in power electronics \cite{arora2018ultrahigh,guo2015tailoring} and field-effect transistors with high temperature electronics are reported. \cite{kim2016exfoliated}
 In bandgap engineering, applying external strain, i.e. bending, starching or compressing the substrates,\cite{rogers2010materials} 
 strain-tunable wavelength selectivity of photodetection, \cite{kang2016crumpled,kawamura2021bandgap} tunable effective mass
 and Schottky barrier height for two-dimensional materials,\cite{lee2011stretchable} are realized in recent years. 
 
 Due to its unusual polarization, two-dimensional III-VI van der Waals (vdW) family members ( i.e. In$_2$Se$_3$, Ga$_2$O$_3$ nanosheets) widely broaden its application ranges to water-splitting\cite{zhao2018two}, gas-sensing,\cite{xie2018functionalization} piezoelectricity,\cite{xue2018multidirection}optoelectronics\cite{jacobs2014extraordinary}and ferroelectrics.\cite{ding2017prediction,xiao2018intrinsic} 
 Cui $et\ al.$ synthesized two-dimensional layered In$_2$Se$_3$ by experiments, and reported its intercorrelated in-plane and out-of-plane ferroelectricity. \cite{cui2018intercorrelated} Reversible gas capture using a ferroelectric switch and molecule multiferroics on the In$_2$Se$_3$ monolayer could be possible.\cite{tang2020reversible} 
 Theoretical calculation demonstrated that two-dimensional gallium and indium oxides is stable via high-throughput computational method,\cite{meng2020two}
 and discovered the near-edge optical properties of $\beta$-Ga$_2$O$_3$ .\cite{mengle2016first}
 High electron mobility and optoelectronic properties of $\beta$-Ga$_2$O$_3$  \cite{guo2019understanding,su2018unusual} make it can be applied in electronics and optoelectronics based on sn-doped Ga$_2$O$_3$ single crystal.\cite{higashiwaki2018guest,usui2018scintillation}
 Wei $et\ al.$ reported the influence of surface vacancies on the electrical and optical properties of $\beta$-Ga$_2$O$_3$.\cite{wei2019modulation} 
 Pearton $et\ al.$ reported applications of field-effect transistors based on $\beta$-Ga$_2$O$_3$.\cite{chabak2016enhancement,pearton2018review}
 And Oh $et\ al.$ investigated solar-blind photodetectors based on Si-implanted $\beta$-Ga$_2$O$_3$\cite{oh2015development}
 
 However, compared with In$_2$Se$_3$ monolayer, the study of electronic and thermal properties on two-dimensional Ga$_2$O$_3$ monolayer still remains the very infant
 step. Quasi-two-dimensional Ga$_2$O$_3$ mentioned in previous literature were obtained by reducing the thickness of bulk Ga$_2$O$_3$, or epitaxial growth of $\beta$-Ga$_2$O$_3$ films prepared by MOCVD.\cite{wang2021effect} .
 Peelaers $et\ al.$ modeled the monolayer Ga$_2$O$_3$ structure by reducing the thickness of the monoclinic $\beta$-Ga$_2$O$_3$.\cite{peelaers2017lack} 
 Atomically controlled surfaces with step and terrace of $\beta$-Ga$_2$O$_3$ single crystal substrates for thin film growth is realized.\cite{ohira2008atomically}
 By plasma etching, the thickness of exfoliated quasi-two-dimensional $\beta$-Ga$_2$O$_3$ flakes can be tuned.\cite{kwon2017tuning}
 Tang $et\ al.$ reported quasi-epitaxial growth of $\beta$-Ga$_2$O$_3$-coated wide band gap semiconductor tape for flexible UV photodetectors.\cite{tang2021quasi}
 Nanomembranes can be exfoliated from $\beta$-Ga$_2$O$_3$  with a thickness less than 100 nm,\cite{kwon2017tuning} and Zhou $et\ al.$ reported its high performance on insulator field-effect transistors.\cite{zhou2017beta}
 Interestingly,  electronic and optoelectronic properties of novel Ga$_2$O$_3$ monolayer can be tuned flexibly,\cite{liao2020tunable}
 which can make applications in gas-sensing devices\cite{zhao2021two} and for high-performance solar blind photodetectors  based on two-dimensional gallium oxide monolayer.\cite{feng2014synthesis}
 Oh $et\ al.$ investigated quasi-two-dimensional $\beta$-gallium oxide solar-blind photodetectors with ultrahigh responsivity.\cite{oh2016quasi}
 Nie $et\ al.$ modulated the blue and green luminescence in the $\beta$-Ga$_2$O$_3$ films\cite{nie2021modulating}
 Swinnich $et\ al.$ studied the electronic applications based on flexible $\beta$-Ga$_2$O$_3$ nanomembrane Schottky barrier diodes.\cite{swinnich2019flexible}
 
 Besides the two-dimensional and bulk Ga$_2$O$_3$, the device based on van der Waals heterostructures widely broaden the applications of these two-dimensional ferroelectric materials.
 Yan $et\ al.$  reported high breakdown electric field in $\beta$-Ga$_2$O$_3$/graphene vertical barrister heterostructure.\cite{yan2018high}
 Tang $et\ al.$ investigated the effect of exciton on Ga$_2$O$_3$/GaN heterostructural ultraviolet photodetectors.\cite{tang2021ga2o3} 
 Li $et\ al.$ investigated deep-ultraviolet photodetection using single-crystalline $\beta$-Ga$_2$O$_3$/NiO heterojunctions.\cite{li2019deep}
 Harada $et\ al.$ reported electric dipole effect in PdCoO$_2$/$\beta$-Ga$_2$O$_3$ Schottky diodes for high-temperature operation.\cite{harada2019electric}
 Kim $et\ al.$ reported ultrahigh deep-UV sensitivity in graphene-gated $\beta$-Ga$_2$O$_3$ phototransistors.\cite{kim2019ultrahigh}
 
 As mentioned by Ding $et\ al.$\cite{ding2017prediction}, the In$_2$Se$_3$ monolayers with quintuple layer(QL) structures could be ferroelectric materials. Here in this study, using the density functional theory, we calculate the polarization
 direction dependent Raman intensity of Raman active modes
 A$^1_1$ and A$^2_1$ of FE-WZ and FE-ZB Ga$_2$O$_3$ monolayer. The vibrations of two major Raman peaks are analyzed. As comparision, the non-resonant Raman spectra of $E^1$ and $E^2$ are also presented.

 \section{Calculation details}
 In this paper, we apply the first-principles density functional theory (DFT) implementing in the Quantum Espresso package, to calculate the optimized crystal structure, electronic properties and phonon band structure.\cite{giannozzi2009quantum}
 The exchange$-$correlation functional is treated with the Perdew–Wang(PW)\cite{perdew1992accurate} local density approximation (LDA), with ultrasoft pseudopotentials applied. 
 The energy cutoff is set as 60 Ry in all calculations. The force
 and electronic convergence tolerance are set to 0.01 eV Å and
 $10^8$ eV, respectively. To minimize the interlayer interactions and make sure the accuracy, the vacuum slab is set to 2.0 nm. A Monkhorst-Pack $\Gamma$-centered k grid is set to 11$\times$11$\times$1 in the structure optimiza-tion, while it is set to 15$\times$15$\times$1 in the self-consistent calculation. What's more,
 the phonon band structures are obtained by diagonalizing
 the force constant matrix with density functional perturbation
 theory (DFPT).\cite{gonze1997dynamical} A Monkhorst-Pack $\Gamma$-centered k grid of 15$\times$15$\times$1 is applied. The force tolerance is set to 10$^{14}$.
 
 \section{Results and discussions}
 \begin{figure}[htbp]
 	\centering
 	\begin{subfigure}[b]{0.95\linewidth}
 		\includegraphics[width=1.0\linewidth]{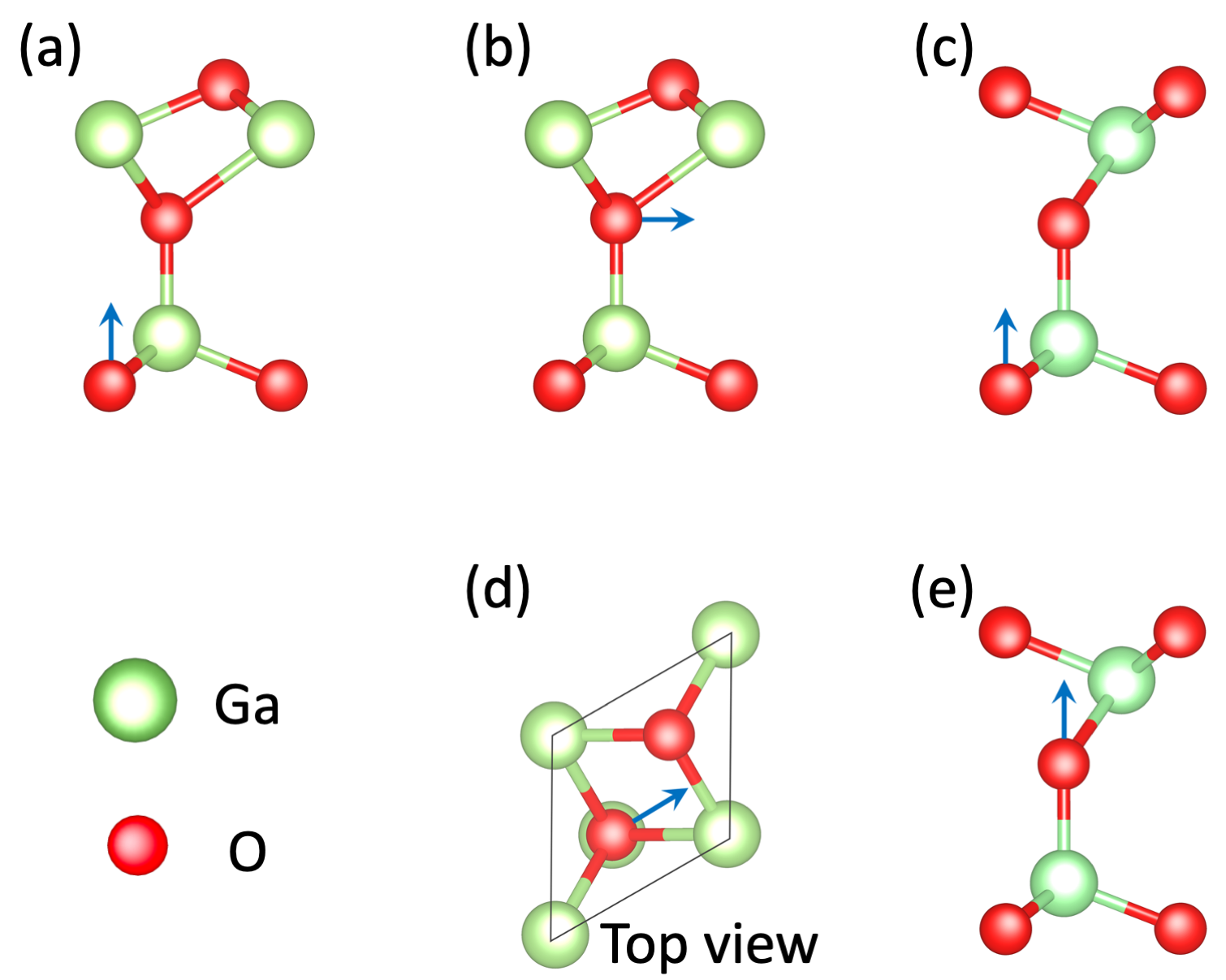}
 	\end{subfigure} 	
 	
 	\caption{\label{fig:intensity} Crystal structure of FE-WZ ((a) for A$^1_1$, (b) for A$^2_1$) and FE-ZB((c) for A$^1_1$, (e) for A$^2_1$) Ga$_2$O$_3$ monolayer, and atomic displacement of active Raman modes. (d) is the top view of (b). The red and green balls are for O and Ga elements, respectively.  The blue arrows refer to the moving vectors of oxygen atoms.}
 	
 \end{figure}

 As discussed in ref.\cite{ding2017prediction}, two kinds of In$_2$Se$_3$ monolayer
 with ferroelectric properties, which can be exfoliated from few-layer In$_2$Se$_3$, were demonstrated stable. As shown in fig.1, the monolayer(QL) contains five atomic layers, and each atomic layer in a QL contains only one elemental species, with the atoms in a given layer arranged in a triangular lattice. According to the different stacking sequence, two ferroelectric structure can be classified as, FE-WZ(ABBAC) and FE-BZ(ABBCA), respectively.
 In this paper, as one of the III-VI van der Waals family members, two phases (FE-WZ and FE-BZ) of Ga$_2$O$_3$ monolayer with QL structure are considered, as shown in fig.1. 
 The five atomic layers in a QL stack in the sequence of O-Ga-O-Ga-O atomic layers.

 We notice that, both FE-WZ and FE-ZB have indirect band structures. As shown in fig.2, for both FE-WZ and FE-ZB, the minimum of conduction band (CBM) locates at $\Gamma $ point, while the maximum of valence band (VBM) locates near K point, which is slightly higher than that at $\Gamma$ point. In the band curve of conduction band, at K point, it is a minimum for FE-WZ Ga$_2$O$_3$ monolayer, while for FE-BZ , it is a maximum.
    \begin{figure}[htbp]
 	\centering
 	\begin{subfigure}[b]{1.0\linewidth}
 		\includegraphics[width=1.0\linewidth]{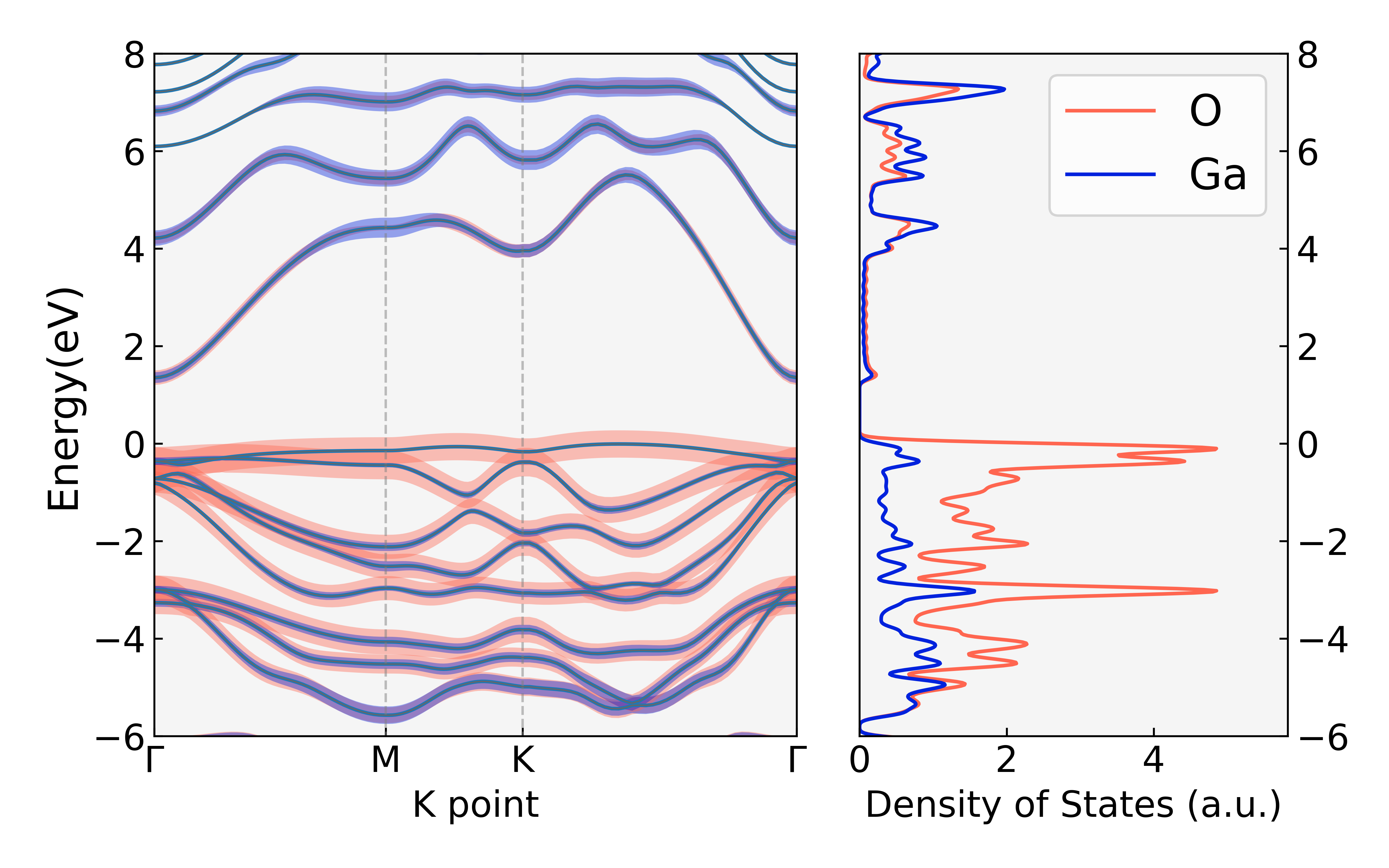}
 		\put(-240,155) {(a)}
 	\end{subfigure} 	
 	\begin{subfigure}[b]{0.99\linewidth}
 		\includegraphics[width=1.0\linewidth]{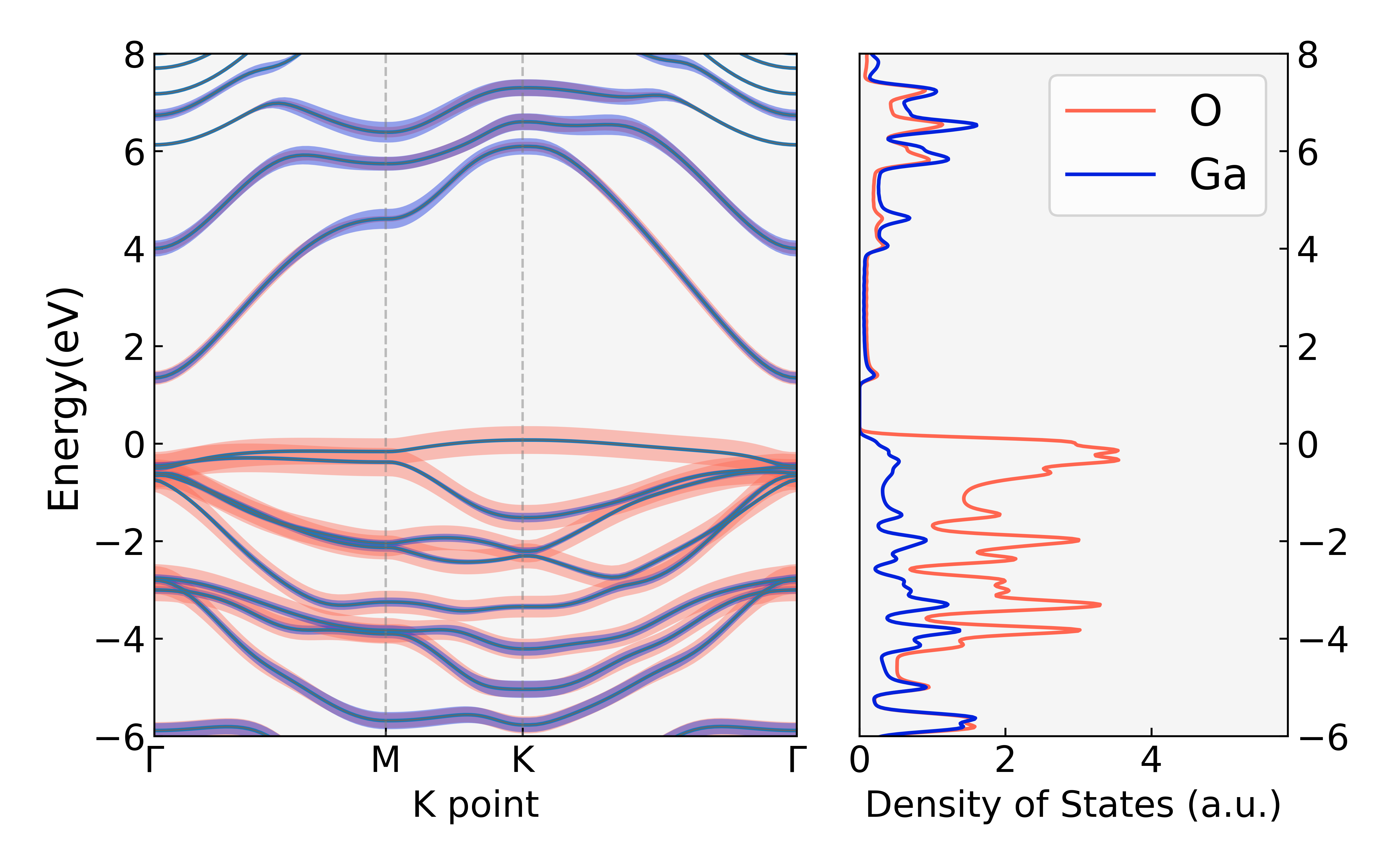}
 		\put(-240,155) {(b)}
 	\end{subfigure} 
 	
 	\caption{\label{fig:intensity} Band structure and PDOS of (a) FE-WZ and (b) FE-ZB Ga$_2$O$_3$ monolayer. The highest point of valence band is set at zero. Atom-projected band structures and PDOS are highlighted with red and blue colors for O and Ga elements, respectively. }	
 \end{figure}

 To better understand the electronic structure of Ga$_2$O$_3$ monolayer, the atom-projected band structures of FE-WZ and FE-ZB are also exhibited in fig.2. We notice that, the main contribution of the valence band comes from oxygen atoms, while for the conduction band, both gallium and oxygen atoms play important roles. This point can also be confirmed from the corresponding partial density of states (PDOS) in the right panel of fig.2. 
 
 The stabilities of Ga$_2$O$_3$ monolayer are essential for the practical device applications.
 By calculating the phonon band structures using DFTP (
 fig.3 a,b), the absence of imaginary phonon modes at $\Gamma$ point indicates that, the crystal structure of monolayer of FE-WZ and FE-BZ could be dynamically stable.
 As shown in the phonon band structure(fig. 3), both FE-WZ and FE-ZB have 12 optical branches(4 one-dimensional modes(A$_1$) and 4 doubly degenerate modes (E) ).

 $$ \Gamma_{\rm optical} = 4{\rm A_1 + 4E} $$
 
  \begin{figure}[htbp]
 	\centering
 	\begin{subfigure}[b]{1.0\linewidth}
 		\includegraphics[width=1.0\linewidth]{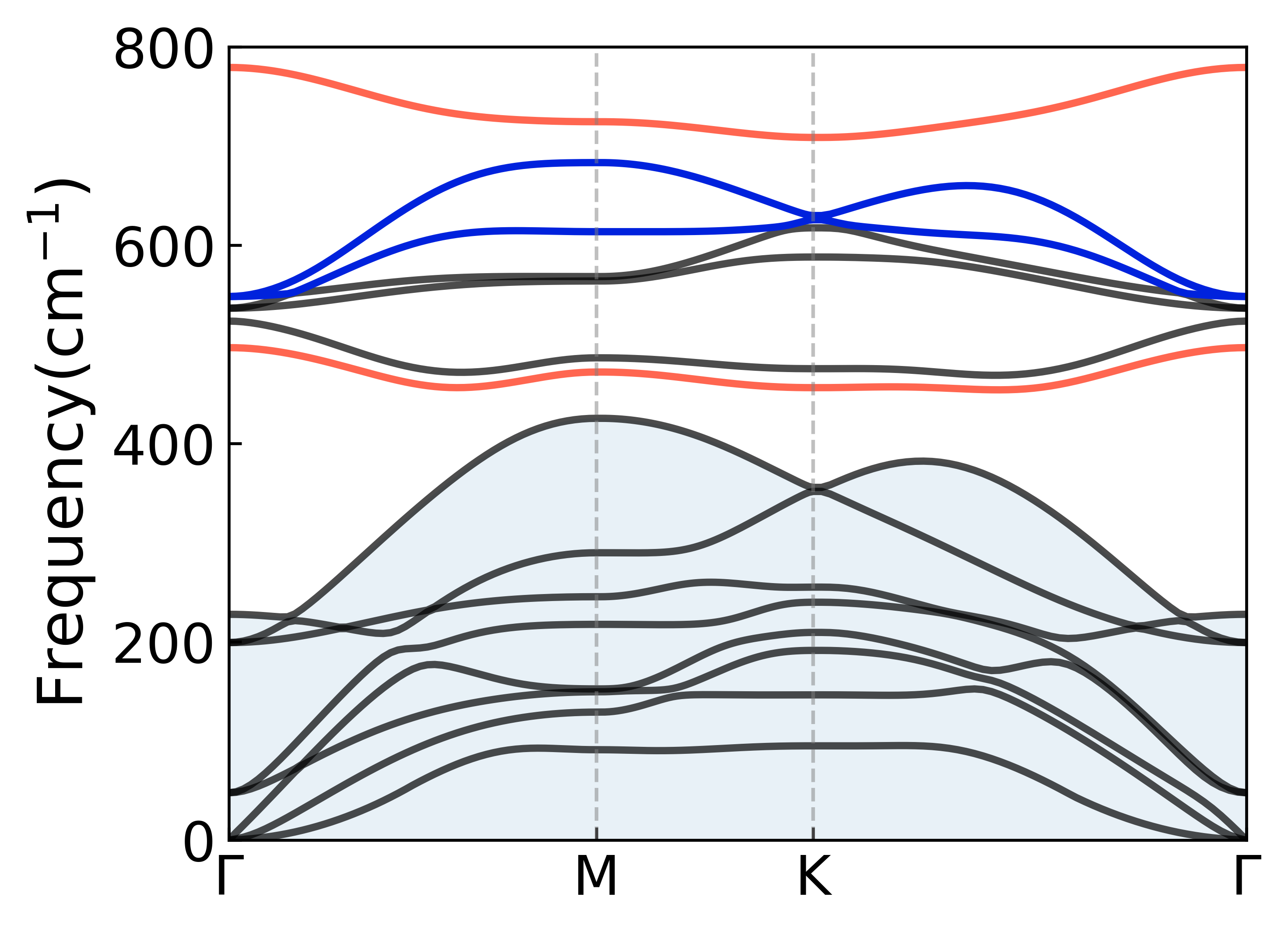}
 		\put(-240,185) {(a)}
 	\end{subfigure} 	
 	\begin{subfigure}[b]{0.99\linewidth}
 		\includegraphics[width=1.0\linewidth]{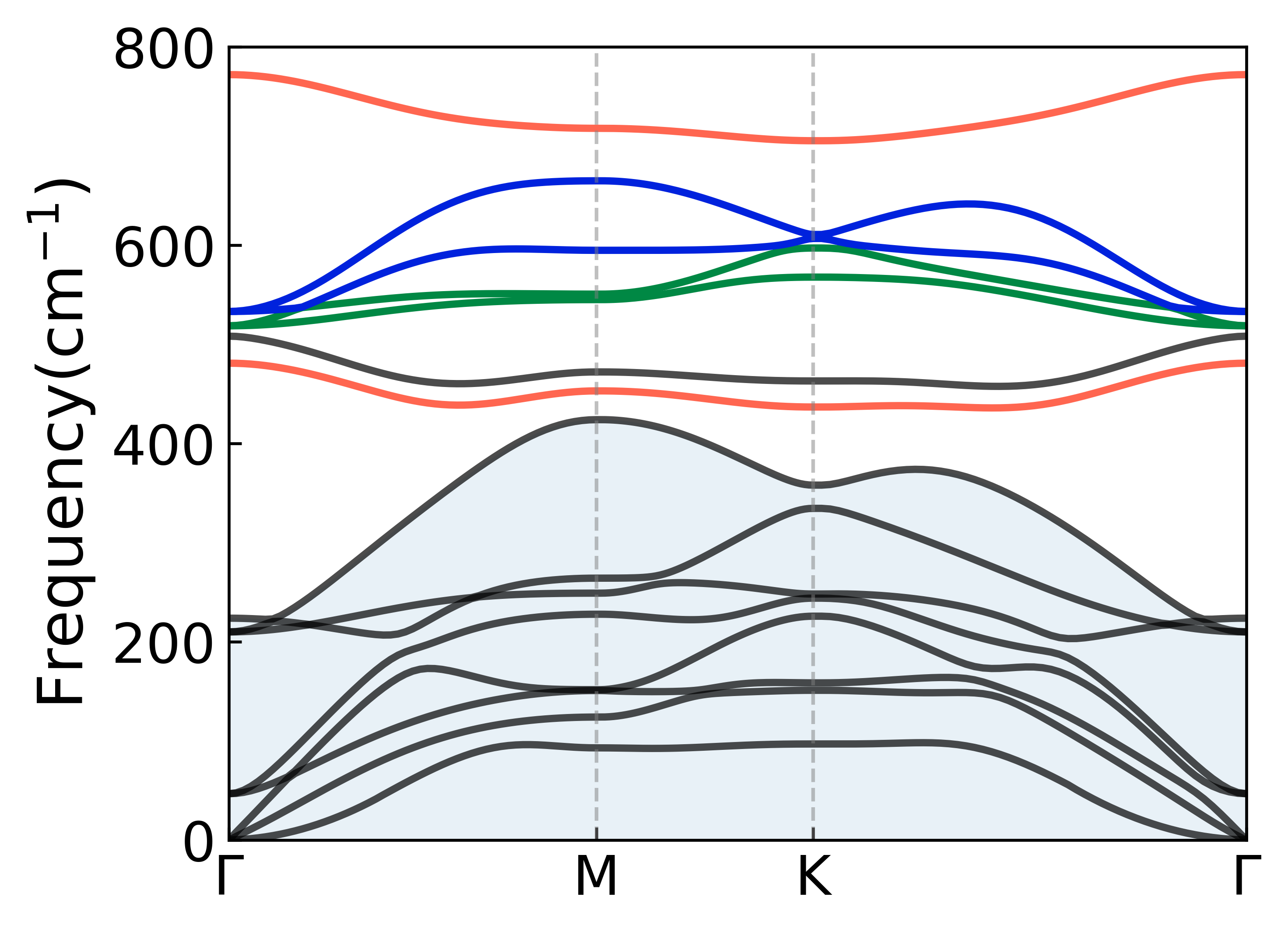}
 		\put(-240,185) {(b)}
 	\end{subfigure} 
 	
 	\caption{\label{fig:phonon} Phonon band structure of (a) FE-WZ and (b) FE-ZB Ga$_2$O$_3$ monolayer along high symmetry k points.  The corresponding Raman active modes (E$^1$(blue), A$_1^1$(red) and A$^2_1$(red)) in
 		FE-WZ and FE-ZB are highlighted by different colors.}	
 \end{figure}

 To better understand which mode is of importance, the non-resonant Raman spectra of FE-WZ and FE-BZ Ga$_2$O$_3$ monolayer are presented in fig.4. The intensities are averaged in all directions of the incident light and all polarization directions of the incident light and scattered light.
 As shown in fig.4(a), FE-WZ Ga$_2$O$_3$ monolayer has only three obvious broadening of Raman peaks( A$_1^1$(487), A$^2_1$(779) and E$^1$(548nm) ). For FE-ZB(fig.4(b)), besides A$_1^1$(481), A$^2_1$(772) and E$^1$(533nm), one more mode is observed E$^2$(519). Compared with the E modes, both A$^1_1$ and A$^2_1$ have much stronger Raman intensity. In the following paragraph, we pay our attention to these two major modes( A$^1_1$ and A$^2_1$ ). In fig.1(a,c), for A$^1_1$ mode, the atomic displacement vibrating along $z$-direction, happens mainly on the oxygen atom located in the lowest layer. For A$^2_1$ mode of FE-WZ(b,d), the atomic displacement vibrating in $x$-$y$ plane, happens mainly on the oxygen atom located in the middle layer. While for A$^2_1$ mode of FE-ZB(c,e), the atomic displacement vibrating in $z$ plane, happens on the oxygen atom located in the middle layer. This demonstrates that, oxygen atoms play an essential role in Raman spectra of Ga$_2$O$_3$ monolayer. 
 
  \begin{figure}[htbp]
 	\centering
 	\begin{subfigure}[b]{1.0\linewidth}
 		\includegraphics[width=1.0\linewidth]{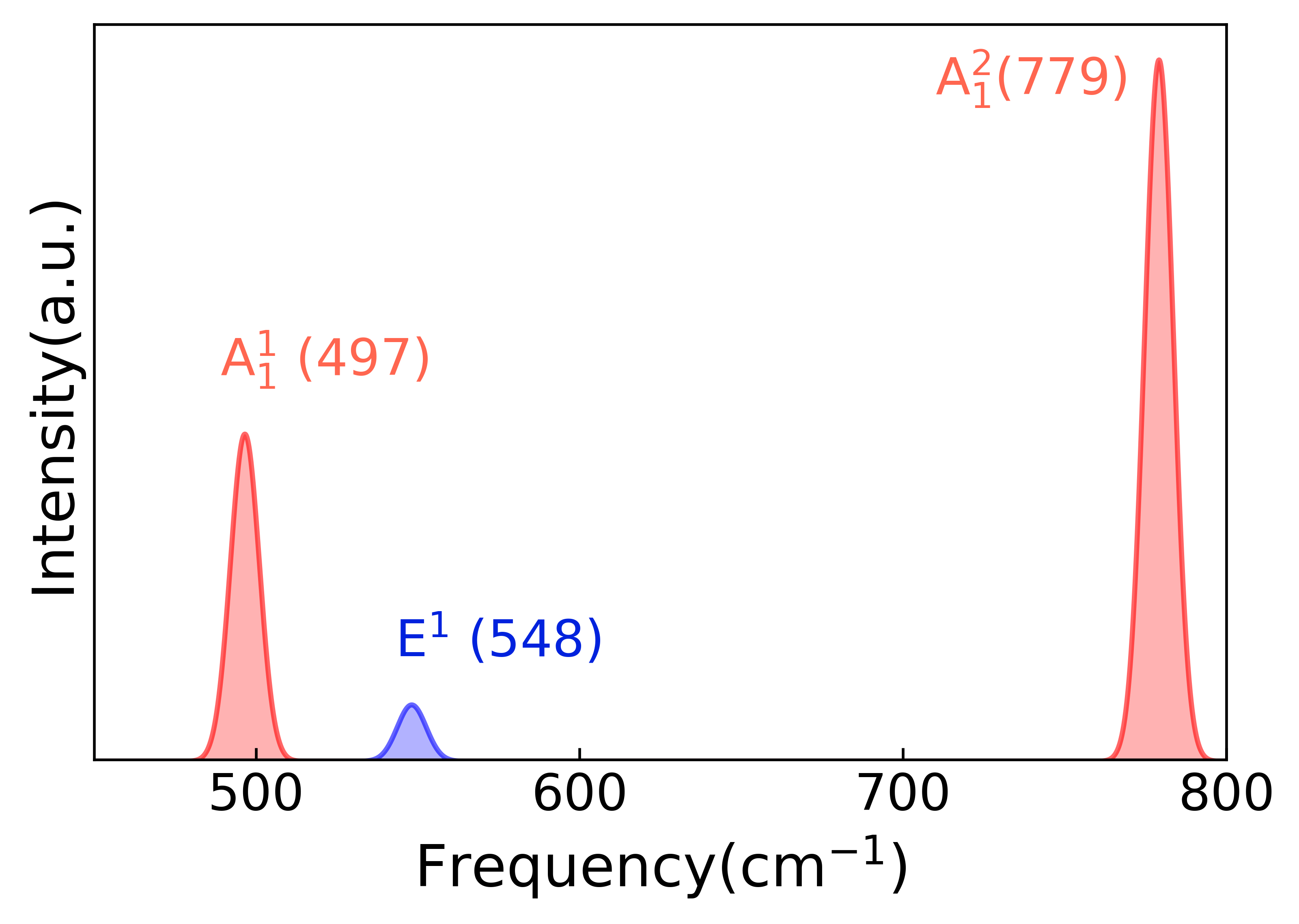}
 		\put(-240,185) {(a)}
 	\end{subfigure} 	
 	\begin{subfigure}[b]{0.99\linewidth}
 		\includegraphics[width=1.0\linewidth]{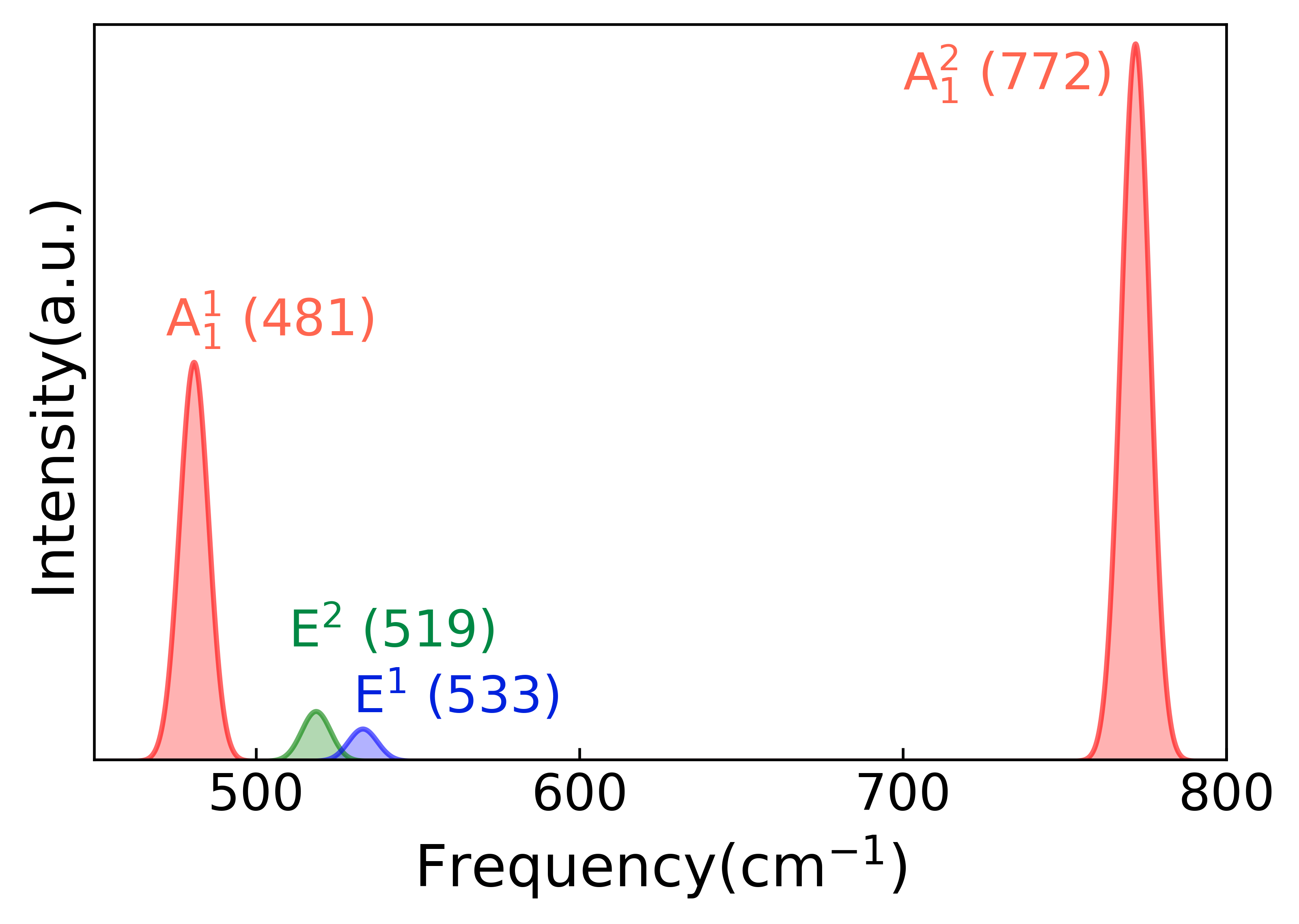}
 		\put(-240,185) {(b)}
 	\end{subfigure} 
 	
 	\caption{\label{fig:intensity} Non-resonant Raman shifts of (a) FE-WZ and (b) FE-ZB Ga$_2$O$_3$ monolayer, with a Gaussian broadening width of 5 cm$^{-1}$.}	
 \end{figure} 
 
 The intensity $I$ of a Raman active mode in a Raman scattering process, can be written as:\cite{umari2001raman,ceriotti2006ab}
 
 $$ I \propto | {\bm e}_{\rm s}\cdot {\bm R} \cdot {\bm e}_{\rm i} |^2. $$
 
 Here, the unitary vectors ${\bm e}_i$ and ${\bm e}_s$ describes the electric field polarization directions of incident and scattered light. Because the Raman tensor $\bf R$ is obtained from the dielectric function, it must be a complex 3 $\times$ 3 matrix for each Raman active modes.
 
 \begin{figure}[htbp]
 	\centering
 	\begin{subfigure}[b]{0.95\linewidth}
 		\includegraphics[width=1.0\linewidth]{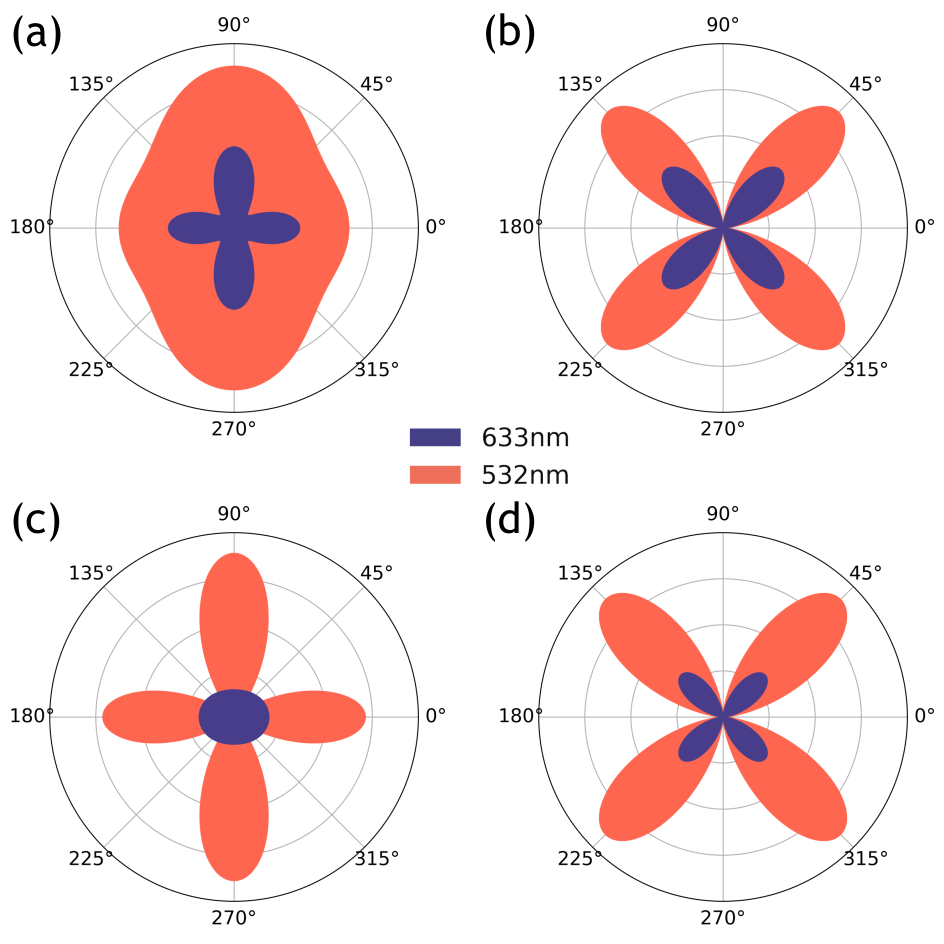}
 	\end{subfigure}

 	\caption{\label{fig:intensity} Polar plots of the incident polarization angle-dependent intensities for A$^1_1$(a,b) and A$^2_1$(c,d) modes of FE-WZ Ga$_2$O$_3$ monolayer, with polarization of scattered light parallel (left panels (a,c)) and perpendicular (right panels (b,d)) to the polarization of the incident light, corresponding to two laser lines (633 (blue) and 532 (red) nm).}	
 \end{figure}

 The matrix element of the Raman tensor $\bm R$ is defined as :
 
 $$
 \bm R_{\alpha\beta}(i) = V_{\rm cell} \sum_{\mu=1}^N\sum_{l=1}^3\frac{\partial \chi_{\alpha\beta}}{\partial r_l(\mu)} \frac{e^i_l(\mu)}{\sqrt{M_{\mu}}},
 $$
 
 Here, $r_l(\mu)$ is the position of the $\mu$th atom along direction $l$, $e^i_l(\mu)$ is the eigen vector of the $i$th phonon mode at the $\Gamma$ point, and $V_{\rm cell}$ is the volume of the unit cell. $\chi_{\alpha\beta}$ is the element of the electric polarizability tensor, which is closely related to the dielectric tensor $\varepsilon_{\alpha\beta}  = 4\pi \chi_{\alpha\beta} + \delta_{\alpha\beta}. $
 The dielectric tensor $\bm \varepsilon$ can be obtained directly from DFT calculations. As the Raman tensor $\bm R$ is complex, we need to consider the influence of light absorption on the Raman spectra. 
 To better understand the impact of light absorption on Raman intensity, two laser lines(633 and 532 nm) are considered in our simulations.

 If the  vector $k$ of the incident light is parallel to the $x$ direction with polarization direction $e_i = ( 0, {\rm cos} \theta, {\rm sin} \theta )$,
 the parallel and perpendicular polarization directions of scattered light can be defined as
 $e^s_{\parallel} = ( 0, {\rm cos} \theta, {\rm sin} \theta )$, $e^s_{\perp} = ( 0, -{\rm sin} \theta, {\rm cos} \theta )$, respectively.
 
 \begin{figure}[htbp]
 	\centering
 	\begin{subfigure}[b]{0.95\linewidth}
 		\includegraphics[width=1.0\linewidth]{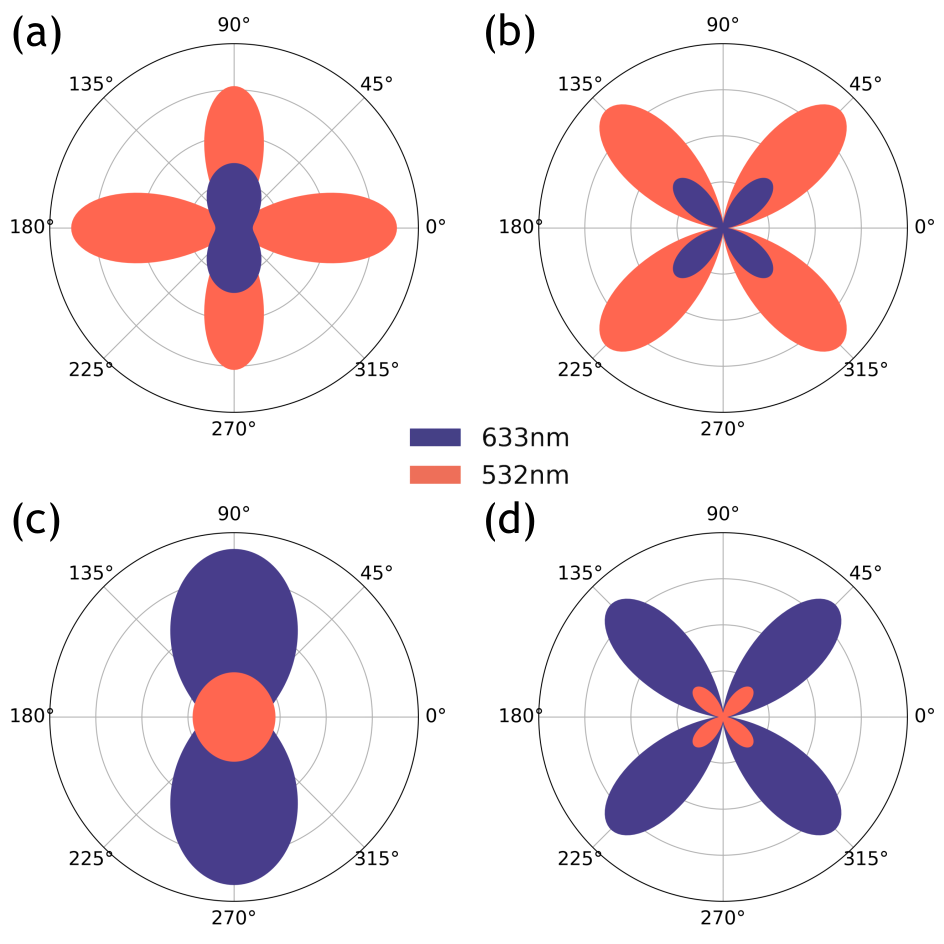}
 	\end{subfigure}

 	\caption{\label{fig:intensity} Polar plots of the incident polarization angle-dependent intensities for A$^1_1$(a,b) and A$^2_1$(c,d) modes of FE-ZB Ga$_2$O$_3$ monolayer, with polarization of scattered light parallel (left panels (a,c)) and perpendicular (right panels (b,d)) to the polarization of the incident light, corresponding to two laser lines (633 (blue) and 532 (red) nm).}	
 \end{figure}
 
 According to the group theory, both FE-WZ and FE-ZB belong to C$_{3v}$ point group.
 The Raman tensor of $A_{1}$ mode is:
 
 \begin{equation*}
 R(A_{1g})=
 \left(
 \begin{array}{ccc}
 a & 0&0 \\
 0&a&0\\
 0&0&b\\
 \end{array}
 \right).
 \end{equation*}

 Here, we take the notations of $a = |a|e^{i\varphi_a}$, and $b = |b|e^{i\varphi_b}$. Herein, for the parallel scattered light, we have:
 $$I^{\parallel}(A_{\rm 1}) \propto |a|^2{\rm cos}^4\theta +|b|^2{\rm sin}^4\theta +\frac{1}{2}|a||b|{\rm sin}^2(2\theta){\rm cos}\varphi_{ab},$$
 and for the perpendicular scattered light:
 $$ I^{\perp}(A_{\rm 1} )\propto \frac{1}{4}\big( |a|^2 +|b|^2 -2|a||b|{\rm cos}\varphi_{ab} \big) {\rm sin}^2(2\theta), $$

 The incident polarization angle-dependent intensities for A$^1_1$(fig.5 a,b) and A$^2_1$(fig.5 c,d) modes of FE-WZ Ga$_2$O$_3$ monolayer are shown in fig. 5, with polarization of scattered light parallel (left panels (a,c)) and perpendicular (right panels (b,d)) to the polarization of the incident light, corresponding to two laser lines (633 (blue) and 532 (red) nm). We notice that, there is no such laser line dependent maxima number change in the angle-dependent Raman intensity of A$^1_1$ mode. When the polarization direction of the scattered light is parallel to the incident light, the four maxima located at $\theta$ = 0$^{\rm o}$, 90$^{\rm o}$, 180$^{\rm o}$ and 270$^{\rm o}$. While for the perpendicular case, they locate at $\theta$ = 45$^{\rm o}$, 135$^{\rm o}$, 225$^{\rm o}$ and 315$^{\rm o}$.
 For mode $A^2_1$,  when the polarization direction of the scattered light is parallel to the incident light, it has two
 maxima for the blue (633 nm) laser line, while four maxima for the red (532 nm) laser line. 
 For the perpendicular case, the configurations of $A^2_1$ is similar to that of $A^1_1$.
 
  \begin{figure}[htbp]
 	\centering
 	\begin{subfigure}[b]{1.0\linewidth}
 		\includegraphics[width=1.0\linewidth]{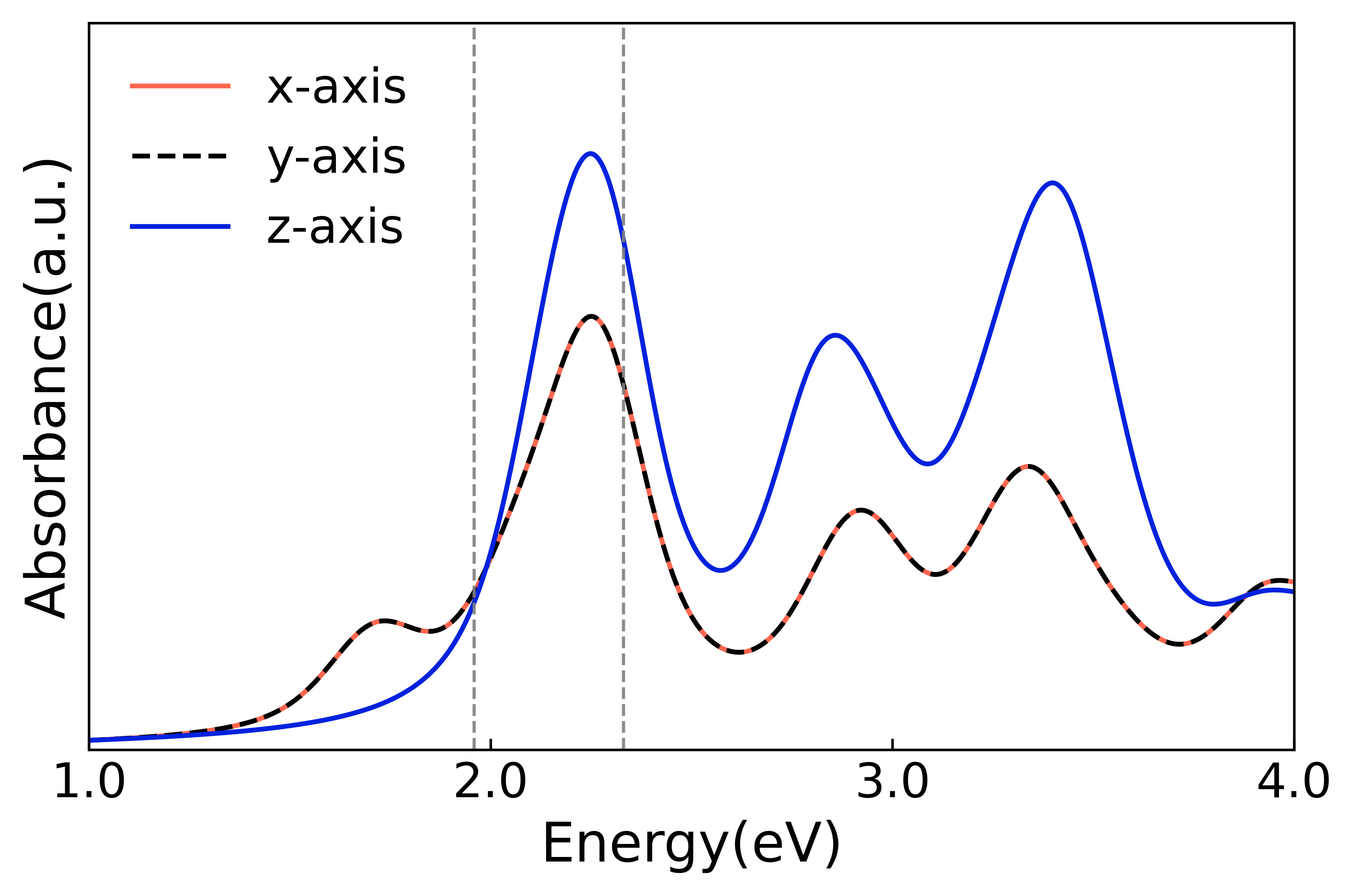}
 		\put(-240,175) {(a)}
 	\end{subfigure} 	
 	\begin{subfigure}[b]{0.99\linewidth}
 		\includegraphics[width=1.0\linewidth]{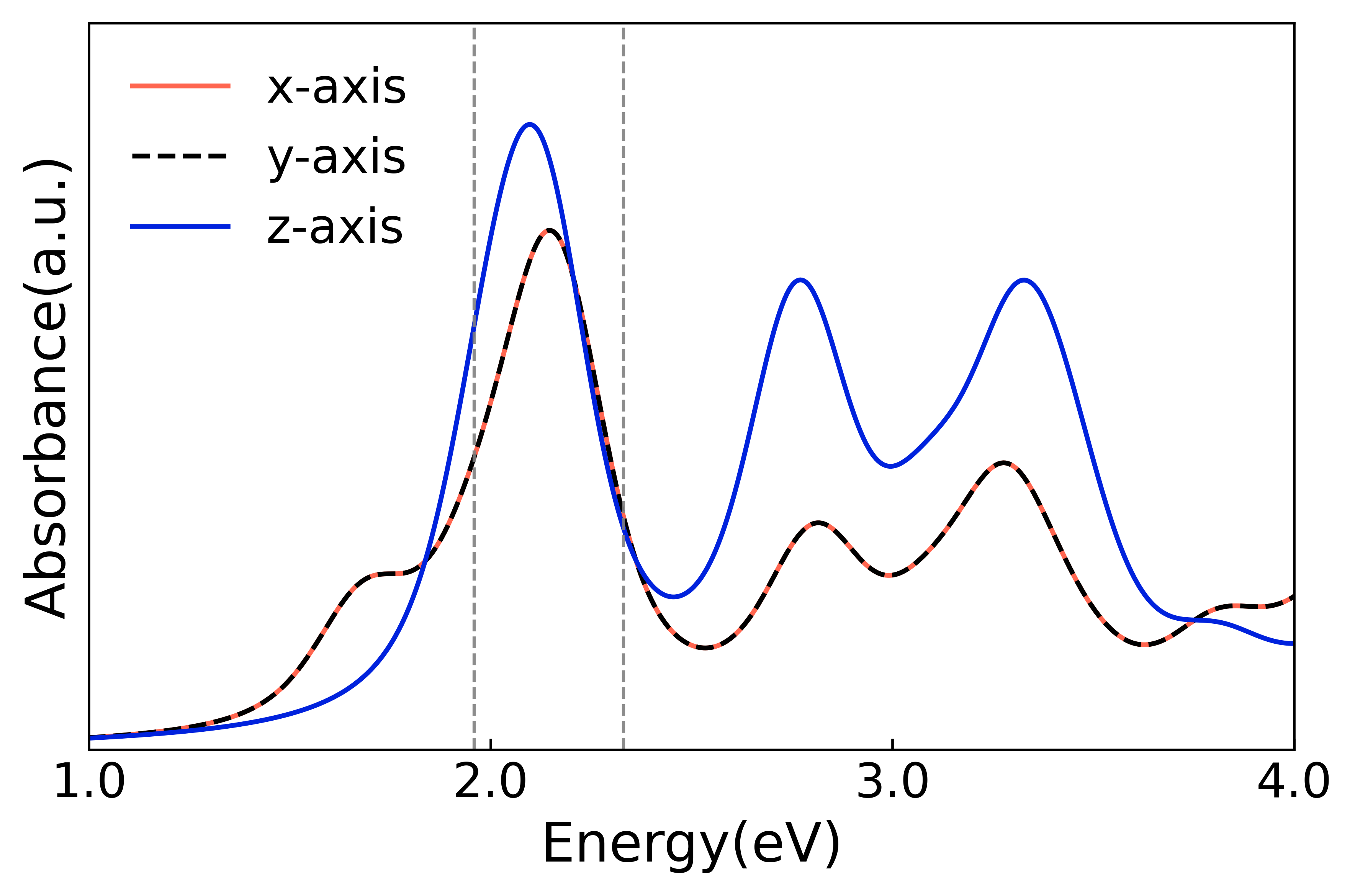}
 		\put(-240,175) {(b)}
 	\end{subfigure} 
 	
 	\caption{\label{fig:intensity} Absorbance spectra of (a) FE-WZ and (b) FE-ZB Ga$_2$O$_3$ monolayer along x, y and z directions. Two vertical dash lines refer the energies of two different lasers( 633nm(1.96eV) and 532nm(2.33eV) ).}	
 \end{figure}
 
 The incident polarization angle-dependent intensities for A$^1_1$(fig.6 a,b) and A$^2_1$(fig.6 c,d) modes of FE-ZB Ga$_2$O$_3$ monolayer are shown in fig. 6, with polarization of scattered light parallel (left panels (a,c)) and perpendicular (right panels (b,d)) to the polarization of the incident light, corresponding to two laser lines (633 (blue) and 532 (red) nm).
 There is no such laser line dependent maxima number change in the angle dependent Raman intensity of A$^1_1$ mode for the red (532nm) laser.
 When the polarization direction of the scattered light is parallel to the incident light, the four maxima located at $\theta$ = 0$^{\rm o}$, 90$^{\rm o}$, 180$^{\rm o}$ and 270$^{\rm o}$. While for the perpendicular case, they locate at $\theta$ = 45$^{\rm o}$, 135$^{\rm o}$, 225$^{\rm o}$ and 315$^{\rm o}$.
 
 However, for the blue (633nm) laser, in the parallel case, it has two
 maxima(90$^0$ and 270$^0$), while it has four maxima in the perpendicular situation. 
 This configuration of $A^2_1$ is similar that of $A^1_1$, eccept the parallel case for the red(532 nm) laser, in which, the configuration of Raman intensity becomes an ellipse.
 
 We notice that the Raman intensity (fig.5) of
 the red (532 nm) laser line is much stronger than that of the
 blue (633 nm) laser line, for both $A_1^1$ and $A^2_1$ modes. 
 This is closely related to the light absorption.  As shown in fig.7, the absorbance at the blue (633 nm) laser line is smaller than that at the red (532 nm) laser line for FE-WZ Ga$_2$O$_3$ monolayer.	Due to the anisotropy of crystal structure, the absorption in x and y direction are almost the identical.

 \section{Conclusions}
 In summary, the vibrational properties and Raman
 spectra of FE-WZ and FE-ZB Ga$_2$O$_3$ monolayers are systematically studied.  The phonon band without imaginary mode demonstrates that both FE-WZ and FE-ZB Ga$_2$O$_3$ monolayers are stable. Besides $E^1$ mode, two major Raman peak ($A^1_1$ and $A^2_1$) are found in both FE-WZ (497, and 779 cm$^{-1}$) and FE-ZB  (481, and 772 cm$^{-1}$)  Ga$_2$O$_3$ monolayers. The angle-dependent Raman intensity of these two major modes are investigated, with polarization of scattered light parallel and perpendicular to the polarizations of the incident light. 
 For FE-WZ Ga$_2$O$_3$ monolayers, the corresponding intensities of laser 532 nm are much stronger than that of laser 633 nm, which is closely related to its absorbance.  
 Therefore, we believe
 that these calculated Raman spectra of FE-WZ and FE-ZB Ga$_2$O$_3$ monolayers should guide rational design of two-dimensional optical devices in future.

 \section*{Conflicts of interest}
 There are no conflicts of interest to declare.
 
 \section*{Acknowledgements}
 This work was supported by the  Project of Improving the Basic Scientific Research Ability of Young and Middle-aged Teachers in Universities of Guangxi(Grant No. 2022KY0784).

  \bibliographystyle{unsrt}
 \bibliography{ref-GaO}
 

 \bigskip 
 
 \bigskip

\end{document}